\documentclass[11pt]{article} 

\newcommand\rf[1]{(\ref{eq:#1})}
\newcommand\lab[1]{\label{eq:#1}}
\newcommand\nonu{\nonumber}
\newcommand\br{\begin{eqnarray}}
\newcommand\er{\end{eqnarray}}
\newcommand\be{\begin{equation}}
\newcommand\ee{\end{equation}}

\newcommand\foot[1]{\footnotemark\footnotetext{#1}}
\newcommand\lb{\lbrack}
\newcommand\rb{\rbrack}

\newcommand\llb{\left\lbrack}
\newcommand\rrb{\right\rbrack}

\renewcommand\({\left(}
\renewcommand\){\right)}
\renewcommand\v{\vert}                     
\newcommand\Bgv{\;\Bigg\vert}            

\newcommand\bc{\begin{center}}
\newcommand\ec{\end{center}}

\newcommand\Tr{\mathop{\mathrm Tr}}                  
\newcommand\partder[2]{\frac{{\partial {#1}}}{{\partial {#2}}}}

\newcommand\Sbr[2]{\Bigl\lbrack\,{#1}\, ,\,{#2}\,\Bigr\rbrack}

\newcommand\Pbr[2]{\Bigl\{ \,{#1}\, ,\,{#2}\,\Bigr\}}  

\renewcommand\a{\alpha}
\renewcommand\b{\beta}

\renewcommand\d{\delta}

\newcommand\vareps{\varepsilon}
\newcommand\g{\gamma}
\newcommand\G{\Gamma}

\newcommand\h{\frac{1}{2}}
\renewcommand\k{\kappa}
\renewcommand\l{\lambda}

\newcommand\m{\mu}
\newcommand\n{\nu}
\renewcommand\o{\over}
\newcommand\om{\omega}
\renewcommand\O{\Omega}

\newcommand\vp{\varphi}
\renewcommand\P{\Phi}
\newcommand\pa{\partial}

\newcommand\pr{\prime}

\newcommand\s{\sigma}

\renewcommand\t{\tau}
\renewcommand\th{\theta}

\newcommand\z{\zeta}

\newcommand\wti{\widetilde}

\newcommand\cA{{\mathcal A}}
\newcommand\cB{{\mathcal B}}
\newcommand\cC{{\mathcal C}}

\newcommand\cE{{\mathcal E}}
\newcommand\cF{{\mathcal F}}

\newcommand\cH{{\mathcal H}}

\newcommand\cL{{\mathcal L}}
\newcommand\cM{{\mathcal M}}
\newcommand\cN{{\mathcal N}}

\newcommand\cP{{\mathcal P}}

\newcommand\cT{{\mathcal T}}

\newcommand{\ct}[1]{\cite{#1}}
\newcommand{\bi}[1]{\bibitem{#1}}
\newcommand\PRD[3]{(#2), \textsl{Phys. Rev.} \textbf{D#1} #3}

\newcommand\PLB[3]{(#2), \textsl{Phys. Lett.} \textbf{#1B} #3}
\newcommand\CQG[3]{(#2), \textsl{Class. Quantum Grav.} \textbf{#1} #3}

\begin{document}

\title{Strings, $p$-Branes and $Dp$-Branes\\ With Dynamical Tension}

\author{Eduardo Guendelman and Alexander Kaganovich\\
\small\it Department of Physics,\\[-1.mm]
\small\it Ben-Gurion University, Beer-Sheva, Israel  \\[-1.mm]
\small\it email: guendel@bgumail.bgu.ac.il, alexk@bgumail.bgu.ac.il \\
${}$ \\
Emil Nissimov and Svetlana Pacheva\\
\small\it Institute for Nuclear Research and Nuclear Energy,\\[-1.mm]
\small\it Bulgarian Academy of Sciences, Sofia, Bulgaria  \\[-1.mm]
\small\it email: nissimov@inrne.bas.bg, svetlana@inrne.bas.bg}
\date{ }
\maketitle

\begin{abstract}
We discuss a new class of brane models (extending both $p$-brane and $Dp$-brane 
cases) where the brane tension appears as an {\em additional dynamical degree of
freedom}~ instead of being put in by hand as an {\em ad hoc}~ dimensionfull scale.
Consistency of dynamics naturally involves the appearence of additional higher-rank
antisymmetric tensor gauge fields on the world-volume 
which can couple to charged
lower-dimensional branes living on the original $Dp$-brane world-volume.
The dynamical tension has the physical meaning of electric-type field strength of
the additional higher-rank world-volume gauge fields. It obeys Maxwell (or 
Yang-Mills) equations of motion (in the string case $p=1$) or their higher-rank
gauge theory analogues (in the $Dp$-brane case). This in particular triggers
a simple classical mechanism of (``color'') charge confinement.    
\end{abstract}    

\section{Introduction}
The crucial relevance of Dirichlet $p$-branes ($Dp$-branes) \ct{Dp-brane-orig},
\textsl{i.e.}, $p+1$-dimensional extended objects in space-time on
which the ends of fundamental open strings can be confined, is largely
appreciated and exploited in modern string theory (for reviews of string and 
brane theories, see \ct{brane-rev,string-brane-books,Dp-brane}). 
Their importance is primarily due to the following basic properties: providing
explicit realization of non-perturbative string dualities, microscopic 
description of black-hole physics, gauge theory/gravity correspondence, 
large-radius compactifications of extra dimensions, brane-world scenarios in 
particle phenomenology, \textit{etc.} .

When building actions in geometrically motivated field theories, which is
the class where strings and branes belong, one of the most important ingredients 
is the consistent generally-covariant integration measure 
density, \textsl{i.e.}, covariant under arbitrary diffeomorphisms 
(reparametrizations) on the underlying space-time manifold. {\em A priori}
there are no compelling geometric reasons restricting us to the natural
choice, which is the standard Riemannian metric density $\sqrt{-g}$ with 
$g \equiv \det\v\v g_{\m\n}\v\v$. For instance, introducing additional $D$ 
scalar fields $\vp^i$ ($i=1,\ldots ,D$ where $D$ is the space-time dimension)
we may employ the following alternative non-Riemannian measure density $\P (\vp)$ :
\be
\P (\vp) \equiv {1\o {D!}}\vareps^{\m_1 \ldots \m_D}
\vareps_{i_1 \ldots i_D} \pa_{\m_1} \vp^{i_1} \ldots \pa_{\m_D} \vp^{i_D} \; .
\lab{D-measure-def}
\ee
Making use of \rf{D-measure-def} allows to construct a broad class of new models
involving Gravity called \textsl{Two-Measure Gravitational Models} \ct{TMT}, whose
actions are typically of the form:
\be
S = \int d^D x\, \P (\vp)\, L_1 + \int d^D x\,\sqrt{-g}\, L_2 \; ,
\lab{TMT-a}
\ee
\be
L_{1,2} = e^{\frac{\a \phi}{M_P}} \Bigl\lb - {1\o \k} R(g,\G ) -
\h g^{\m\n}\pa_\m \phi \pa_\n \phi
+ \bigl(\mathrm{Higgs}\bigr) + \bigl(\mathrm{fermions}\bigr)\Bigr\rb \; .
\lab{TMT-b}
\ee
Here $R(g,\G )$ is the scalar curvature in the first-order formalism
(\textsl{i.e.}, the connection $\G$ is independent of the metric),
$\phi$ is the dilaton field, $M_P$ is the Planck mass, \textsl{etc.}. 
Although naively the additional ``measure-density'' scalars $\vp^i$ appear in 
\rf{TMT-a} as pure-gauge degrees of freedom (due to the invariance under arbitrary
diffeomorphisms in the $\vp^i$-target space), there is still a remnant -- the so
called ``geometric'' field $\chi (x) \equiv \frac{\P (\vp)}{\sqrt{-g}}$, which 
remains an additional dynamical degree of freedom beyond the standard physical 
degrees of freedom characteristic to the ordinary gravity models with the standard
Riemannian-metric integration measure. The most important property of the 
``geometric'' field $\chi (x)$ is that its dynamics is determined solely through the
matter fields locally (\textsl{i.e.}, without gravitational interaction).
The latter turns out to have a significant impact on the physical properties
of the two-measure gravity models which allows them to address various basic
problems in cosmology and particle physics phenomenology and provide
physically plausible solutions, for instance: (i) the issue of scale invariance
and its dynamical breakdown, \textsl{i.e.}, spontaneous generation of
dimensionfull fundamental scales; (ii) cosmological constant problem;  
(iii) geometric origin of fermionic families.  

In what follows we are going to apply the above ideas to
the case of string, $p$-brane and $Dp$-brane models. To make the exposition
self-contained Sections 2 and 3 below review our earlier works 
\ct{mstring-orig,mstring} by providing a detailed description of the simplest
$p=1$ case - the class of new modified-measure string models. Then in Section 4
we proceed with the extension of our construction to the physically most
interesting case of new modified-measure $Dp$-brane models. Along the way we
elaborate on various important properties of the modified-measure string and brane
models with dynamical string/brane tension.

\section{Bosonic Strings with a Modified World-Sheet Integration Measure}
First let us recall the standard Polyakov-type action for the bosonic string 
\ct{Polyakov}:
\be
S_{\mathrm{Pol}} = - T \int d^2\s\,
\h\sqrt{-\g}\g^{ab}\pa_a X^\m \pa_b X^\n G_{\m\n}(X) \; .
\lab{string-action-Pol}
\ee
Here $(\s^0,\s^1) \equiv (\t,\s)$; $a,b =0,1$; $\m,\n = 0,1,\ldots ,D-1$;
$G_{\m\n}$ denotes the Riemannian metric on the embedding space-time;
$\g_{ab}$ is the intrinsic Riemannian metric on the $1+1$-dimensional string 
world-sheet and $\g = \det\v\v\g_{ab}\v\v$; $T$ indicates the string tension 
-- a dimensionfull scale introduced \textsl{ad hoc}.
The resulting equations of motion w.r.t. $\g^{ab}$ and $X^\m$ read,
respectively:
\br
T_{ab}\equiv \(\pa_a X^\m \pa_b X^\n - 
\h \g_{ab} \g^{cd} \pa_c X^\m \pa_d X^\n\) G_{\m\n}(X) = 0 \; ,
\lab{gamma-eqs-Pol} \\
{1\o {\sqrt{-\g}}}\pa_a \(\sqrt{-\g}\g^{ab}\pa_b X^\m\) +
\g^{ab}\pa_a X^\n \pa_b X^\l \G^\m_{\n\l} = 0 \; ,
\lab{X-eqs-Pol}
\er
where $\G^\m_{\n\l}=\h G^{\m\k}\(\pa_\n G_{\k\l}+\pa_\l G_{\k\n}-\pa_\k G_{\n\l}\)$ 
is the connection for the external metric.

Now, following the same ideas as in the construction of two-measure gravity
theories \rf{TMT-a}--\rf{TMT-b} let us introduce two additional world-sheet 
scalar fields $\vp^i$ ($i=1,2$) and replace $\sqrt{-\g}$ with a new 
reparametrization-covariant world-sheet integration measure density $\P(\vp)$
defined solely in terms of $\vp^i$:
\be
\P (\vp) \equiv \h \vareps_{ij} \vareps^{ab} \pa_a \vp^i \pa_b \vp^j
= \vareps_{ij} {\dot{\vp}}^i \pa_\s \vp^j \; .
\lab{def-measure}
\ee
However, the naively generalized string action:
\br
S_{1} = -\h \int d^2\s\,\P (\vp) \g^{ab} \pa_a X^\m \pa_b X^\n G_{\m\n}(X)
\nonu
\er
has a problem: the equations of motion w.r.t. $\g^{ab}$ lead to an unacceptable
condition $\P (\vp)\,\pa_a X^\m \pa_b X^\n G_{\m\n}(X) = 0$, \textsl{i.e.},
vanishing of the induced metric on the world-sheet.

To remedy the above situation let us consider topological (total derivative) terms
$\int \O$ w.r.t. the standard Riemannian world-sheet integration measure. 
Upon measure replacement $\sqrt{-\g} \to \P (\vp)$, \textsl{i.e.},
$\int \O\ \to \int \O\,\frac{\P (\vp)}{\sqrt{-\g}}$, the former are 
{\em not any more} topological -- they will contribute nontrivially to the 
equations of motion. For instance:
\be
\int d^2\s\,\sqrt{-\g}\, R  \to \int d^2\s\,\P (\vp) \, R \quad ,\;\;
R = \frac{\vareps^{ab}}{2\sqrt{-\g}} \(\pa_a \om_b - \pa_b \om_a\) \; ,
\lab{Euler-mod}
\ee
where $R$ is the scalar curvature w.r.t. $D=2$ spin-connection
$\om_{a}^{\bar{a}\bar{b}}=\om_{a}\vareps^{\bar{a}\bar{b}}$ (here $\bar{a},\bar{b}$
denote tangent space indices). Note that the vector field $\om_{a}$ behaves as 
world-sheet Abelian gauge field.

Eq.\rf{Euler-mod} prompts us to construct the following consistent modified
bosonic string action\foot{In ref.\ct{K-V} another interesting geometric
modification of the standard bosonic string model has been proposed, which
is based on dynamical world-sheet metric and torsion.}:
\be
S = - \int d^2\s \,\P (\vp) \Bigl\lb \h \g^{ab} \pa_a X^{\m} \pa_b X^{\n} G_{\m\n}
- \frac{\vareps^{ab}}{2\sqrt{-\g}} F_{ab}(A)\Bigr\rb \; ,
\lab{string-action}
\ee
where $\P (\vp)$ is given by \rf{def-measure} and 
$F_{ab}(A) \equiv \pa_{a} A_{b} - \pa_{b} A_{a}$ is the field-strength of an
auxiliary Abelian gauge field $A_a$. The action \rf{string-action} is
reparametrization-invariant as its ordinary string analogue \rf{string-action-Pol}. 
Furthermore, \rf{string-action} is invariant under diffeomorphisms in 
$\vp$-target space supplemented with a special conformal transformation of 
$\g_{ab}$ : 
\be
\vp^{i} \longrightarrow \vp^{\pr\, i} = \vp^{\pr\, i} (\vp) \quad ,\quad
\g_{ab} \longrightarrow \g^{\pr}_{ab} = J \g_{ab} \;\; ,\;\;
J \equiv \det \Bigl\Vert \frac{\pa\vp^{\pr\, i}}{\pa\vp^j} \Bigr\Vert \; .
\lab{Phi-Weyl-symm}
\ee
The latter symmetry, which we will call {\em ``$\P$-extended Weyl symmetry''},
is the counterpart of the ordinary Weyl conformal symmetry of the standard
string action \rf{string-action-Pol}.

The equations of motion of the action \rf{string-action} w.r.t. $\vp^i$ :
\be
\vareps^{ab} \pa_{b} \vp^{i} \pa_{a} \Bigl(
\g^{cd} \pa_{c} X^{\m}\pa_{d} X^{\n} G_{\m\n}(X) -
\frac {\vareps^{cd}}{\sqrt{-\g}} F_{cd} \Bigr) = 0
\lab{vp-eqs}
\ee
imply (provided $\P (\vp) \neq 0$) :
\be
\g^{cd} \pa_{c} X^{\m}\pa_{d} X^{\n} G_{\m\n}(X) -
\frac {\vareps^{cd}}{\sqrt{-\g}} F_{cd} = M \; \Bigl( = \mathrm{const}\Bigr)
\; .
\lab{vp-eqs-a}
\ee
The equations of motion w.r.t. $\g^{ab}$ are:
\be
T_{ab} \equiv \pa_{a} X^{\m}\pa_{b} X^{\n} G_{\m\n}(X)   
- \h\g_{ab} \frac{\vareps^{cd}}{\sqrt{-\g}}F_{cd} = 0 \; .
\lab{gamma-eqs}
\ee
Both Eqs.\rf{vp-eqs-a}--\rf{gamma-eqs} yield $M=0$ and: 
\be
\Bigl( \pa_{a} X^{\m}\pa_{b} X^{\n} - 
\h \g_{ab} \g^{cd} \pa_{c} X^{\m}\pa_{d} X^{\n} \Bigr) G_{\m\n}(X) = 0 \; ,
\lab{Pol-eqs}
\ee
which is the same as in standard Polyakov-type formulation \rf{gamma-eqs-Pol}.

The equations of motion w.r.t. $X^\m$ read:
\be
\pa_a \(\P \g^{ab}\pa_b X^\m\) +
\P \g^{ab}\pa_a X^\n \pa_b X^\l \G^\m_{\n\l} = 0  \; ,
\lab{X-eqs}
\ee
where again $\G^\m_{\n\l}$ is the connection corresponding to the 
external space-time metric $G_{\m\n}$ as in the standard string case 
\rf{X-eqs-Pol}.

Now, let us consider the equations of motion w.r.t. $A_a$ resulting from 
\rf{string-action} :
\be
\vareps^{ab} \pa_{b} \Bigl(\frac{\P (\vp)}{\sqrt{-\g}}\Bigr) = 0 \; .
\lab{A-eqs}
\ee
The latter can be integrated to yield a {\em spontaneously induced} string tension: 
\br
\frac {\P (\vp)}{\sqrt{-\gamma}} = \textrm{const} \equiv T \; .
\nonu
\er

Since the modified-measure string model \rf{string-action} naturally
requires the presence of the auxiliary Abelian world-sheet gauge field $A_a$,
we may extend it by introducing a coupling of $A_a$ to some world-sheet
charge current $j^a$ :
\be
S = - \int d^2\s \,\P (\vp) \Bigl\lb \h \g^{ab} \pa_a X^{\m} \pa_b X^{\n} G_{\m\n}
- \frac{\vareps^{ab}}{2\sqrt{-\g}} F_{ab}(A)\Bigr\rb 
+ \int d^2\s \, A_a j^a \; .
\lab{string-action-plus}
\ee
In particular, we may take $j^a$ to be the current of point-like charges on
the string, so that in the ``static'' gauge:
\be
\int d^2\s \, A_a j^a =  - \sum_i e_{i} \int d\t A_0 (\t,\s_{i}) \; ,
\lab{static-gauge-term}
\ee
where $\s_i$ ($0 < \s_1 <\ldots < \s_N \leq 2\pi$) are the locations of the
charges. Now, instead of \rf{A-eqs} the action \rf{string-action-plus} produces 
the following $A_a$-equations of motion:
\be
\vareps^{ab}\pa_b E + j^a = 0 \quad ,\quad E \equiv \frac{\P (\vp)}{\sqrt{-\g}}
\; .
\lab{A-eqs-a}
\ee
Note that Eqs.\rf{A-eqs-a} look exactly as $D=2$ Maxwell equations where the 
{\em variable} dynamical string tension $E\equiv \P (\vp)/\sqrt{-\gamma}$ is
identified as world-sheet electric field strength, \textsl{i.e.}, canonically
conjugated momentum w.r.t. $A_1$.

The physical meaning of the dynamical string tension as world-sheet electric
field strength can be directly verified in the framework of the canonical 
Hamiltonian treatment of the modified-measure string model \rf{string-action-plus}.
Indeed, from the explicit form of the action \rf{string-action-plus} we find
the canonical momenta to be:
\br
\pi^{\vp}_i = - \vareps_{ij} \pa_\s \vp^j 
\Bigl\lb \h \g^{ab} \pa_a X^{\m} \pa_b X^{\n} G_{\m\n}
- \frac{\vareps^{ab}}{2\sqrt{-\g}} F_{ab}(A)\Bigr\rb \; ,
\lab{can-mom-a} \\
\pi_{A_1} \equiv E = \frac{\P (\vp)}{\sqrt{-\g}} \;\; ,\;\;
\cP_\m = - \P (\vp) \(\g^{00}\dot{X}^\n + \g^{01}\pa_\s X^\n\) G_{\m\n} \; ,
\lab{can-mom-b}
\er
Using \rf{can-mom-a}--\rf{can-mom-b} we obtain the canonical Hamiltonian as a 
linear combination of first-class constraints only. Part of the latter resemble
the constraints in the ordinary string case $\pi_{\g^{ab}}=0 $ and
\br
\cT_{\pm} \equiv \frac{1}{4} G^{\m\n}
\Bigl(\frac{\cP_\m}{E} \pm G_{\m\k} \pa_\s X^\k \Bigr)
\Bigl(\frac{\cP_\n}{E} \pm G_{\n\l} \pa_\s X^\l \Bigr) = 0 \; ,
\nonu
\er
where in the last Virasoro constraints the dynamical string tension $E$
appears instead of the {\em ad hoc} constant tension.

The rest of the Hamiltonian constraints are $\pi_{A_0} = 0$ and
\be
\pa_\s E - \sum_i e_{i} \d (\s - \s_{i}) = 0 \; ,
\lab{Gauss-law}
\ee
which is precisely the $D=2$ ``Gauss law'' constraint for the dynamical string
tension coinciding with the $0$-th component of the Maxwell-type Eq.\rf{A-eqs-a}.
Finally, we have constraints involving only the measure-density fields:
\be
\pa_\s \vp^i \pi^{\vp}_i = 0 \quad ,\quad
\frac{\pi^{\vp}_2}{\pa_\s \vp^1} = 0  \; .
\lab{vp-constr}
\ee
The last two constraints span a closed Poisson-bracket algebra:
\br
\Pbr{\pa_\s \vp^i \pi^{\vp}_i (\s)}{\pa_{\s^\pr} \vp^i \pi^{\vp}_i (\s^\pr)} =
\phantom{aaaaaaaaaa}
\nonu \\
2 \pa_\s \vp^i \pi^{\vp}_i (\s) \pa_\s \d (\s -\s^\pr) + 
\pa_\s \(\pa_\s \vp^i \pi^{\vp}_i\) \d (\s -\s^\pr) \; ,
\nonu
\er
(a centerless Virasoro algebra), and:
\br
\Pbr{\pa_\s \vp^i \pi^{\vp}_i (\s)}{\frac{\pi^{\vp}_2}{\pa_\s \vp^1}(\s^\pr)} =
- \pa_\s \Bigl(\frac{\pi^{\vp}_2}{\pa_\s \vp^1}\Bigr) \d (\s -\s^\pr) \; .
\nonu
\er
Therefore, the constraints \rf{vp-constr} imply that the measure-density 
scalars $\vp^i$ are pure-gauge degrees of freedom. The only physical remnant
of the latter is the specific combination $\frac{\P (\vp)}{\sqrt{-\g}}$ 
(see first Eq.\rf{can-mom-b}) -- the world-sheet electric field-strength
simultaneously playing the role of dynamical string tension.

\section{Non-Abelian Generalization}

We will now show that it is possible to introduce an alternative form of
topological term necessary to make consistent the modified-measure string
model which is built in terms of a \textit{non-Abelian} auxiliary world-sheet
gauge field.

First, let us notice the following identity in $D=2$ involving Abelian gauge 
field $A_a$:
\be
{1\o {2\sqrt{-\g}}}\vareps^{ab} F_{ab}(A) = 
\sqrt{\h F_{ab}(A) F_{cd}(A) \g^{ac}\g^{bd}}  \; .
\lab{D2-ident}
\ee
This suggests the following proper extension of the modified-measure bosonic string
action \rf{string-action} involving a non-Abelian (\textsl{e.g.},
$SU(\cN)$) auxiliary gauge field $A_a$ (here we take for simplicity flat 
external metric $G_{\m\n}=\eta_{\m\n}$) :
\br
S = - \int d^2\s \,\P (\vp) \Bigl\lb \h \g^{ab} \pa_a X^{\m} \pa_b X_\m 
-\sqrt{\h \Tr (F_{ab}(A)F_{cd}(A)) \g^{ac}\g^{bd}}\Bigr\rb
\nonu  \\
= - \int d^2\s \,\P (\vp) \Bigl\lb \h \g^{ab} \pa_a X^{\m} \pa_b X_\m 
- \frac{1}{\sqrt{-\g}} \sqrt{\Tr (F_{01}(A)F_{01}(A))}\Bigr\rb \; ,
\lab{string-action-NA}
\er
where $F_{ab}(A) = \pa_a A_b - \pa_b A_c + i \bigl\lb A_a,\, A_b\bigr\rb$.

The action \rf{string-action-NA} is again invariant under the {\em $\P$-extended Weyl
(conformal)} symmetry \rf{Phi-Weyl-symm}.

Notice that the {\em ``square-root'' Yang-Mills}~ action (with the regular
Rie\-mann\-ian metric integration measure):
\be
\int d^2\s\, \sqrt{-\g} \sqrt{\h \Tr (F_{ab}(A) F_{cd}(A)) \g^{ac}\g^{bd}}
= \int d^2\s\, \sqrt{\Tr (F_{01}(A) F_{01}(A))}
\lab{sqrt-YM}
\ee
is a ``topological'' action similarly to the $D=3$ Chern-Simmons action,
\textsl{i.e.}, it is \textit{metric-independent}.

Similarly to the Abelian case \rf{string-action-plus} we can also add a coupling of 
the auxiliary non-Abelian gauge field $A_a$ to an external ``color''-charge
world-sheet current $j^a$:
\br
S = - \int d^2\s \,\P (\vp) \Bigl\lb \h \g^{ab} \pa_a X^{\m} \pa_b X_\m
- \frac{1}{\sqrt{-\g}} \sqrt{\Tr (F_{01}(A)F_{01}(A))}\Bigr\rb
\nonu \\
+ \int d^2\s \, \Tr\( A_a j^a\) \; .  \phantom{aaaaaaaaaaaaaaaaaaaaaa}
\lab{string-action-NA-plus}
\er
In particular, for a current of ``color'' point-like charges on the
world-sheet in the ``static'' gauge :
\be
\int d^2\s \,\Tr\( A_a j^a\) = - \sum_{i}\Tr C_{i}\int d\t A_{0}(\t ,\s_{i}) \; ,
\lab{static-gauge-term-NA}
\ee
where $\s_i$ ($0 < \s_1 <\ldots < \s_N \leq 2\pi$) are the locations of the
charges.

The action \rf{string-action-NA-plus} produces the following equations of 
motion w.r.t. $\vp^i$ and $\g^{ab}$, respectively:
\be
\h \g^{cd}\pa_c X^\m \pa_d X_\m -\frac{1}{\sqrt{-\g}}
\sqrt{\Tr (F_{01} F_{01})} = M \; \Bigl( = \mathrm{const}\Bigr) \; ,
\lab{vp-eqs-NA}
\ee
\be
T_{ab} \equiv 
\pa_a X^\m \pa_b X_\m -\frac{1}{\sqrt{-\g}}\g_{ab}\sqrt{\Tr (F_{01} F_{01})}=0
\; .
\lab{gamma-eqs-NA}
\ee
As in the Abelian case the above Eqs.\rf{vp-eqs-NA}--\rf{gamma-eqs-NA} imply 
$M=0$ and the Polyakov-type equation \rf{gamma-eqs-Pol}.

Similarly to the Abelian case \rf{A-eqs-a}, the equations of motion
of \rf{string-action-NA-plus} w.r.t. the auxiliary gauge field $A_a$
resemble the $D=2$ non-Abelian Yang-Mills equations:
\be
\vareps^{ab} \nabla_b \cE + j^a = 0  \; ,
\lab{A-eqs-NA}
\ee
where:
\be
\nabla_a \cE \equiv \pa_a \cE + i \bigl\lb A_a,\,\cE \bigr\rb  \quad ,\quad
\cE \equiv \pi_{A_1} \equiv \frac{\P (\vp)}{\sqrt{-\g}}
\frac{F_{01}}{\sqrt{\Tr (F_{01} F_{01})}} \; .
\lab{E-def-NA}
\ee
Here $\cE$ is the non-Abelian electric field-strength -- the canonically 
conjugated momentum $\pi_{A_1}$ of $A_1$, whose norm is the dynamical string
tension $T \equiv |\cE| = \P (\vp)/\sqrt{-\g}$.

The equations of motion for the dynamical string tension following from 
\rf{A-eqs-NA} is:
\be
\pa_a \Bigl(\frac{\P (\vp)}{\sqrt{-\g}}\Bigr) + 
\vareps_{ab} \frac{\Tr\( F_{01} j^b\)}{\sqrt{\Tr\( F_{01}^2\)}} = 0 \; .
\lab{Tension-eqs}
\ee
In particular, in the absence of external charges ($j^a =0$) :
$T \equiv \P (\vp)/\sqrt{-\g} = T_0 \equiv \mathrm{const}$

Finally, the $X^\m$-equations of motion $\pa_a \(\P (\vp) \g^{ab} \pa_b X^\m\)=0$
resulting from the action \rf{string-action-NA-plus} can be rewritten in the 
conformal gauge $\sqrt{-\g} \g^{ab} = \eta^{ab}$ as:
\be
\frac{\P (\vp)}{\sqrt{-\g}} \pa^a \pa_a X^\m  - {\wti j}^a \pa_a X^\m = 0
\quad ,\;\; \mathrm{where} \;\;
{\wti j}_a \equiv 
\vareps_{ab}\frac{\Tr\( F_{01} j^b\)}{\sqrt{\Tr\( F_{01}^2\)}} \; .
\lab{X-eqs-NA}
\ee

For static charges ${\wti j}_1 = - \sum_i {\wti e}_i \d (\s -\s_i)$ :
\be
T \equiv \P (\vp)/\sqrt{-\g} = T_0 + \sum_i {\wti e}_i \th (\s -\s_i) 
\quad ,\quad
{\wti e}_i \equiv \frac{\Tr\( F_{01} C_i\)}{\sqrt{\Tr\( F_{01}^2\)}}\Bgv_{\s =\s_i}
\; ;
\lab{Tension-sol}
\ee
\be
T \pa^a \pa_a X^\m + \Bigl(\sum_i {\wti e}_i \d (\s -\s_i)\Bigr)\pa_\s X^\m = 0
\;\; \to
\left\{\begin{array}{ll}
\pa^a \pa_a X^\m = 0 \\
\pa_\s X^\m \Bgv_{\s =\s_i} = 0
\end{array} \right. \; .
\lab{X-sol}
\ee


Let us return to the $D=2$ Yang-Mills-like Eqs.\rf{A-eqs-NA} whose $0$-th 
component $\pa_\s \cE + i\Sbr{A_1}{\cE} + j^0 = 0$ is the ``Gauss law'' constraint
for the dynamical string tension ($T \equiv |\cE| = \P (\vp)/\sqrt{-\g}$).
For point-like ``color'' charges and using the gauge $A_1 =0$ (\textsl{i.e.},
$\cE \to {\wti \cE} = G \cE G^{-1}$ where $A_1 = -i G^{-1} \pa_\s G$),
the latter reads:
\be
\pa_\s {\wti \cE} -\sum_{i} {\wti C}_{i} \d (\s - \s_{i}) = 0 \quad ,\quad
{\wti C}_{i} \equiv G C_i G^{-1} \Bgv_{\s =\s_i} \; .
\lab{Gauss-law-NA}
\ee

Let us consider the case of {\em closed} modified string with positions of the
``color'' charges at $0 <\s_1 <\ldots <\s_N \leq 2\pi$. Then, integrating the
``Gauss law'' constraint \rf{Gauss-law-NA} along the string (at fixed proper
time) we obtain:
\be
\sum_i {\wti C}_i = 0 \quad ,\quad {\wti \cE}_{i,i+1} = {\wti \cE}_{i-1,i} + 
{\wti C}_i \; ,
\lab{charge-constr-NA}
\ee
where ${\wti \cE}_{i,i+1} = {\wti \cE}$ in the interval $\s_i < \s <\s_{i+1}$. 

The implications of Eqs.\rf{Tension-sol}--\rf{charge-constr-NA} can be
summarize as follows:
\begin{itemize}
\item
Eqs.\rf{Tension-sol}--\rf{X-sol} tell us that the modified-measure
(closed) string with $N$ point-like (``color'') charges on it 
(\rf{string-action-plus} or \rf{string-action-NA-plus}) is equivalent to $N$ 
chain-wise connected regular open string segments (stretching from $\s_i$ to
$\s_{i+1}$, $i=0,1,\ldots ,N-1$) which obey Neumann boundary conditions. 
\item
Each of the above open string segments, with end-points at the charges
$e_{i}$ and $e_{i+1}$ (in the Abelian case) or $C_{i}$ and $C_{i+1}$ (in the 
non-Abelian case), has {\em different} constant string tension $T_{i,i+1}$
such that $T_{i,i+1} = T_{i-1,i} + \stackrel{(\sim )}{e}_i$ (the non-Abelian
${\wti e}_i$ are defined in \rf{Tension-sol}).
\item
Eq.\rf{charge-constr-NA} shows that the only (classically) admissable 
configuration of ``color'' point-like charges coupled to a modified-measure 
closed bosonic string is the one with {\em zero} total ``color'' charge, 
\textsl{i.e.}, the model \rf{string-action-NA-plus} provides a classical 
mechanism of ``color'' charge confinement.
\end{itemize}

\section{$Dp$-Branes With Dynamical Tension}

Our construction from the previous two sections can be extended to the case
of higher-dimensional extended objects with dynamical tension embracing both
(ordinary) $p$-branes and $Dp$-branes. First, let us recall the standard 
formulation of $Dp$-branes given in terms of the Dirac-Born-Infeld (DBI) action
\ct{Dp-brane} :
\be
S_{DBI} = -T\int d^{p+1}\s\,\llb e^{-\a U}\sqrt{-\det\vert\vert G_{ab} - \cF_{ab} \vert\vert}
+ \frac{\vareps^{a_1 \ldots a_{p+1}}}{p+1}\cC_{a_1 \ldots a_{p+1}}\rrb
\; ,
\lab{DBI-action}
\ee
with the following short-hand notations:
\be
G_{ab} \equiv \pa_a X^\m \pa_b X^\n G_{\m\n}(X) \quad ,\quad
\cF_{ab} \equiv \pa_a X^\m \pa_b X^\n B_{\m\n}(X) - F_{ab}(A)\; ,
\lab{G-F-notations}
\ee
\be
\cC_{a_1 \ldots a_{p+1}} \equiv 
\pa_{a_1} X^{\m_1} \ldots \pa_{a_{p+1}} X^{\m_{p+1}} C_{\m_1\ldots\m_{p+1}}(X)
\; .
\lab{C-notation}
\ee
Here $G_{\m\n}(X)$,  $U(X)$, $B_{\m\n}(X)$, and $C_{\m_1\ldots\m_{p+1}}(X)$ are the 
background metric, the dilaton, $2$-form Neveu-Schwarz and $(p+1)$-form 
Ramond-Ramond gauge fields, respectively, whereas 
$F_{ab}(A) = \pa_a A_b - \pa_b A_b$ is the 
field-strength of the Abelian world-volume gauge field $A_a$. All
world-volume indices take values $a,b=0,1,\ldots ,p$ and 
$\vareps^{a_1 \ldots a_{p+1}}$ is the $(p+1)$-dimensional totally
antisymmetric tensor ($\vareps^{0 1\ldots p} = 1$).

Similarly to the string case we now introduce a modified world-volume
integration measure density in terms of $p+1$ auxiliary scalar fields $\vp^i$
($i=1,\ldots ,p+1$) :
\be
\P (\vp) \equiv \frac{1}{(p+1)!} \vareps_{i_1\ldots i_{p+1}} 
\vareps^{a_1\ldots a_{p+1}} \pa_{a_1} \vp^{i_1}\ldots \pa_{a_{p+1}} 
\vp^{i_{p+1}} \; ,
\lab{brane-measure}
\ee
and use it to construct the following new $p$-brane-type action (coupling to
the Ramond-Ramond background field is omitted for simplicity)\foot{Some time
ago a Polyakov-type action classically equivalent to the original DBI-action
of the $Dp$-brane \rf{DBI-action} has been proposed in \ct{Hull-Zeid} :
$S = - T \int d^{p+1}\s\,\sqrt{-\z}
\Bigl\lb e^{-\b U} \h\z^{ab}\( G_{ba} - \cF_{ba}\) - (p-1)\Bigr\rb$. 
The advantage of the present more general action \rf{mDBI-action-0} is
that not only it yields variable dynamical brane tension and naturally
introduces additional higher-rank world-volume gauge fields, but also it does
not need a ``cosmological'' term.}:
\be
S = - \int d^{p+1}\s\, \P (\vp) \Bigl\lb e^{-\b U} \h\z^{ab}\( G_{ba} - \cF_{ba}\)
+ {1\o \sqrt{-\z}} \O (\cA)\Bigr\rb + \int d^{p+1}\s\,\cL (\cA)  \; .
\lab{mDBI-action-0}
\ee
Here apart from \rf{G-F-notations} the following new notations
are used. The $(p+1)\times (p+1)$ matrix $\z_{ab}$ of auxiliary variables is
an arbitrary world-volume $2$-tensor, $\z^{ab}$ denotes the corresponding
inverse matrix ($\z^{ac} \z_{cb} = \d^a_b$) and $\z \equiv \det\v\v\z_{ab}\v\v$. 
The term $\O (\cA)$ indicates a topological density given in terms of some 
additional auxiliary gauge (or matter) fields $\cA^I$ living on the world-volume,
where ``topological'' means:
\be
\partder{\O}{\cA^I} - \pa_a \(\partder{\O}{\pa_a \cA^I}\) = 0 \;\;
\mathrm{identically} \quad ,\quad
\mathrm{i.e.}\;\; \d\O (\cA) = \pa_a \(\partder{\O}{\pa_a \cA^I} \d \cA^I\) \; . 
\lab{top-density-def}
\ee
$\cL (\cA)$ describes possible coupling of the auxiliary fields $A^I$ to
external ``currents'' on the brane world-volume.

The requirement for $\O (\cA)$ to be a topological density is dictated by the
requirement that the new brane action \rf{mDBI-action-0} (in the 
absence of the last gauge/matter term $\int d^{p+1}\s\,\cL (\cA)$) reproduces 
the standard $Dp$-brane equations of motion resulting from the DBI action
\rf{DBI-action} apart from the fact that the $Dp$-brane tension 
$T \equiv \P (\vp)/\sqrt{-\z}$ becomes now an {\em additional dynamical degree
of freedom} (note that no \textit{ad hoc} dimensionfull tension factor $T$ has
been introduced in \rf{mDBI-action-0}). 
Let us stress that, similarly to the string case, the modified-measure
brane model \rf{mDBI-action-0} naturally requires (through the necessity to
introduce topological density $\O (\cA)$) the existence on the
world-volume of an additional (higher-rank tensor) gauge field $\cA$ apart
from the standard world-volume Abelian vector gauge $A_a$.

Splitting the auxiliary tensor variable $\z^{ab} = \g^{ab} + \z^{[ab]}$ into
symmetric and anti-symmetric parts and setting $\z^{[ab]}=0$, the action
\rf{mDBI-action-0} reduces to the action of the modified-measure model of ordinary
$p$-branes \ct{mstring} with Neveu-Schwarz field $B_{\m\n}$ and
world-volume gauge field $A_a$ disappearing and $\g_{ab}$ assuming the role of 
world-volume Riemannian metric.

The most obvious example of a topological density $\O (\cA)$ for the
additional auxiliary gauge/matter world-volume fields in \rf{mDBI-action-0} is:
\be
\O (\cA) = -\frac{\vareps^{a_1\ldots a_{p+1}}}{p+1} F_{a_1\ldots a_{p+1}} (\cA)
\quad ,\quad
F_{a_1\ldots a_{p+1}} (\cA) = (p+1)\pa_{\lb a_1} \cA_{a_2\ldots a_{p+1}\rb} \; ,
\lab{top-density-p-rank}
\ee
where $\cA_{a_1 \ldots a_p}$ denotes rank $p$~ antisymmetric tensor (Abelian) 
gauge field on the world-volume. Further, as a physically interesting example
let us consider the following natural coupling of the auxiliary $p$-form gauge
field:
\be
\int d^{p+1}\s\,\cL (\cA) = 
\int d^{p+1}\s\, \cA_{a_1\ldots a_p} j^{a_1\ldots a_p}
\lab{brane-A-coupling}
\ee
to an external world-volume current:
\be
j^{a_1\ldots a_p} = \sum_i e_i 
\int_{\cB_i} d^p u\, \frac{1}{p!} \vareps^{\a_1\ldots\a_p}
\partder{\s_i^{a_1}}{u^{\a_1}}\ldots \partder{\s_i^{a_p}}{u^{\a_p}}
\d^{(p+1)} \bigl(\underline{\s} - \underline{\s}_i (\underline{u})\bigr) \; .
\lab{p-form-current}
\ee
Here $j^{a_1\ldots a_p}$ is a current of charged $(p-1)$-sub-branes $\cB_i$ 
embedded into the original $p$-brane world-volume via
$\s^a = \s^a_i (\underline{u})$ with parameters
$\underline{u} \equiv (u^\a)_{\a =0,\ldots, p-1}$. For simplicity we assume
that the $\cB_i$ sub-branes do not intersect each other. 

With the choices \rf{top-density-p-rank} and 
\rf{brane-A-coupling}--\rf{p-form-current} the action \rf{mDBI-action-0} becomes:
\br
S = - \int d^{p+1}\s\, \P (\vp) \Bigl\lb e^{-\b U} \h\z^{ab} \( G_{ba} - \cF_{ba}\)
- \frac{\vareps^{a_1 \ldots a_{p+1}}}{(p+1)\sqrt{-\z}} 
F_{a_1 \ldots a_{p+1}} (\cA) \Bigr\rb   
\nonu \\ 
- \int d^{p+1}\s\,\cA_{a_1\ldots a_p}
\vareps^{a a_1\ldots a_p} {\wti j}_a \; , \phantom{aaaaaaaaaaaaaa}
\lab{mDBI-action-1}
\er
with ${\wti j}_a = \sum_i e_i \cN^{(i)}_a$,
where $\cN^{(i)}_a$ is the normal vector w.r.t. world-hyper\-sur\-face of the 
$(p-1)$-sub-brane $\cB_i$ :
\be
\cN^{(i)}_a \equiv {1\o {p!}} \vareps_{a b_1\ldots b_p}
\int_{\cB_i} d^p u\, \frac{1}{p!} \vareps^{\a_1\ldots\a_p}
\partder{\s_i^{a_1}}{u^{\a_1}}\ldots \partder{\s_i^{a_p}}{u^{\a_p}}
\d^{(p+1)} \bigl(\underline{\s} - \underline{\s}_i (\underline{u})\bigr)
\; .
\lab{def-N}
\ee

Let us note that some time ago a modified $p$-brane model has been proposed in 
ref.\ct{Bergshoeff-etal} which also contains world-volume $p$-form gauge field
$\cA_{a_1 \ldots a_p}$. However, the latter model is significantly different 
from \rf{mDBI-action-1} since it is not of Polyakov-type and, moreover,
$\cA_{a_1 \ldots a_p}$ appears there quadratically rather than linearly.

More generally, for $p+1=rs$ we can have a more general type of topological
density entering \rf{mDBI-action-0} :
\be
\O (\cA) = {1\o {rs}}\vareps^{a_{11}\ldots a_{1r} \ldots a_{s1}\ldots a_{sr}} 
F_{a_{11}\ldots a_{1r}}(\cA) \ldots F_{a_{s1}\ldots a_{sr}}(\cA)
\lab{top-density-rs}
\ee
with rank $r-1$ (smaller than $p$) auxiliary world-volume gauge fields.

We may also employ {\em non-Abelian}~ auxiliary world-volume gauge fields as in 
the string case. For instance, when $p=3$ we may take:
\be
\O (\cA) = {1\o 4}\vareps^{abcd} \Tr\( F_{ab}(\cA) F_{cd}(\cA)\)
\lab{top-density-NA}
\ee
or, more generally, for $p+1=2q$ :
\be
\O (\cA) = 
{1\o {2q}}\vareps^{a_1 b_1\ldots a_q b_q} 
\Tr\( F_{a_1 b_1}(\cA)\ldots F_{a_q b_q}(\cA) \) \; ,
\lab{top-density-NA-q}
\ee
where $F_{ab}(\cA) = \pa_a\cA_b - \pa_b\cA_a + i \bigl\lb \cA_a,\,\cA_b\bigr\rb$.

The modified-measure brane action \rf{mDBI-action-1} yields the following
equations of motion w.r.t. $\vp^i$ and $\z^{ab}$ : 
\be
e^{-\b U} \h \z^{ab} \( G_{ba} - \cF_{ba}\) + {1\o \sqrt{-\z}} \O (\cA) = 
M \equiv \mathrm{const} \; ,
\lab{brane-vp-eqs}
\ee
\be
e^{-\b U} \( G_{ab} - \cF_{ab}\) + \z_{ab} \frac{1}{\sqrt{-\z}} \O (\cA) = 0 \; .
\lab{z-eqs}
\ee
Both Eqs.\rf{brane-vp-eqs}--\rf{z-eqs} imply:
\be
\z^{ab} \( G_{ba} - \cF_{ba}\) = 2M \frac{p+1}{p-1} e^{\b U} \quad ,\quad
\frac{1}{\sqrt{-\z}} \O (\cA) = - \frac{2M}{p-1}
\lab{brane-trace-eqs}
\ee
which when substituted in \rf{z-eqs} give:
\be
G_{ab} - \cF_{ab} = \frac{2M}{p-1} e^{\b U}\,\z_{ab}
\lab{G-F-z-eqs}
\ee

We now consider the equations of motion of the modified brane action \rf{mDBI-action-0} 
w.r.t. auxiliary (gauge) fields $\cA^I$ -- these are the equations
determining the dynamical brane tension $\cT \equiv \P (\vp)/\sqrt{-\z}$ :
\be
\pa_a \Bigl(\frac{\P (\vp)}{\sqrt{-\z}}\Bigr)\,\partder{\O}{\pa_a \cA^I} + j_I = 0
\; ,
\lab{brane-A-eqs}
\ee
where 
$j_I \equiv \partder{\cL}{\cA^I} - \pa_a \Bigl(\partder{\cL}{\pa_a \cA^I}\Bigr)$
is the corresponding ``current'' coupled to $\cA^I$. Note that the
topological nature of $\O (\cA)$ (Eq.\rf{top-density-def}) is crucial in the
derivation of \rf{brane-A-eqs}. With the choice
\rf{brane-A-coupling}--\rf{p-form-current}, Eq.\rf{brane-A-eqs}  becomes:
\be
\pa_a \Bigl(\frac{\P (\vp)}{\sqrt{-\z}}\Bigr) + {\wti j}_a = 0 \quad ,\quad
{\wti j}_a \equiv \sum_i e_i \cN^{(i)}_a \; .
\lab{brane-tension-eq}
\ee
It is straighforward to deduce from the action \rf{mDBI-action-1} that, similarly
to the modified string case \rf{A-eqs-a}, the dynamical brane tension is 
identified as the electric-type field strength, \textsl{i.e.}, the canonical 
momentum corresponding to the $p$-form gauge field:
$\pi_{A_{1\ldots p}} \equiv E = \frac{\P (\vp)}{\sqrt{-\z}}$. Accordingly,
Eqs.\rf{brane-tension-eq} are of the same form as the Maxwell-type equations of 
motion for the $p$-form gauge field in $p+1$ dimensions.

In particular, in the absence of coupling of external world-volume currents
to the auxiliary (gauge) fields $\cA^I$ Eq.\rf{brane-A-eqs} 
(or \rf{brane-tension-eq}) imply:
\be
\cT \equiv \P (\vp)/\sqrt{-\z} = C \equiv \mathrm{const}
\lab{brane-tension-const}
\ee

Now, using Eqs.\rf{brane-vp-eqs} and \rf{G-F-z-eqs} it is straightforward to 
show that the modified brane action \rf{mDBI-action-0} with $\cL (\cA)=0$
classically reduces to the standard $Dp$-brane DBI-action \rf{DBI-action} :
\be
S^\pr_{DBI} = -T^\pr \int d^{p+1}\s\, e^{-\b^\pr U}
\sqrt{-\det\vert\vert G_{ab} - \cF_{ab} \vert\vert}  \; ,
\lab{DBI-action-a}
\ee
\be
T^\pr \equiv \h C (2M)^{-\frac{p-1}{2}} (p-1)^{\frac{p+1}{2}} \quad ,\quad 
\b^\pr \equiv \frac{p+1}{2}\b \; ,
\lab{brane-tension-const-a}
\ee
where, however, the $Dp$-brane tension $T^{\pr}$ is dynamically generated
according to \rf{brane-tension-const} and \rf{brane-tension-const-a}.

More generally, recalling the definition \rf{def-N} of $\cN^{(i)}_a$
we find from \rf{brane-tension-eq} that the dynamical brane tension
$\cT\equiv \P (\vp)/\sqrt{-\z}$ is piece-wise constant on the world-volume with 
jumps when crossing the world-hypersurface of each charged $(p-1)$-sub-brane 
$\cB_i$. The corresponding jump being equal to the charge magnitude $\pm e_i$ 
(the overall sign depending on the direction of crossing w.r.t. the normal
$\cN^{(i)}_a$).

Finally, let us consider the equations of motion of \rf{mDBI-action-0} w.r.t. the 
Abelian $Dp$-brane vector gauge field $A_a$ :
\be
\pa_b \(\P (\vp) \z^{[ab]} e^{-\b U}\) = 0  \; ,
\lab{brane-A-vec-eqs}
\ee
and the equations of motion w.r.t. the embedding coordinates $X^\m$ (taking
into account Eqs.\rf{brane-A-vec-eqs}) :
\br
\pa_a \(\P (\vp) \z^{(ab)} \pa_b X^\m \) 
- \h \P (\vp) G^{\m\n} \b \pa_\n U \,\z^{[ab]} F_{ab}(A) +
\nonu \\
\P (\vp) \pa_a X^\l \pa_b X^\n \Bigl\lb \z^{(ab)} \G^\m_{\l\n}({\bar G})
+ \h \z^{[ab]} G^{\m\k} \Bigl(\b \pa_\k U B_{\l\n}
- \cH_{\k\l\n}(B) \Bigr)\Bigr\rb = 0  \; .
\lab{brane-X-eqs}
\er
where $\G^\m_{\l\n}({\bar G})$ is the connection of the rescaled
background metric ${\bar G}_{\m\n} = e^{-\b U} G_{\m\n}$ :
\be
\G^\m_{\l\n}({\bar G}) =\h {\bar G}^{\m\k}
\(\pa_\l {\bar G}_{\k\n}+\pa_\n {\bar G}_{\k\l}-\pa_\k {\bar G}_{\l\n}\)
\lab{affine-def}
\ee
and $\cH_{\k\l\n}({\bar B})$ is the field-strength of the background 
Neveu-Schwarz two-form gauge field: 
\be
\cH_{\k\l\n}(B) = \pa_\k B_{\l\n} + \pa_\n B_{\k\l} + \pa_\l B_{\n\k}
\lab{NS-curv}
\ee
Using \rf{brane-tension-eq}, both Eqs.\rf{brane-A-vec-eqs}--\rf{brane-X-eqs}
can be rewritten in the form:
\be
\frac{\P (\vp)}{\sqrt{-\z}}\pa_b {\wti H}^{ab}
- \sum_i e_i \cN^{(i)}_b {\wti H}^{ab} = 0   \quad ,\quad
{\wti H}^{ab} \equiv \sqrt{-\z}\,\z^{[ab]} e^{-\b U}  \; ,
\lab{brane-A-vec-eqs-a}
\ee
and:
\be
\frac{\P (\vp)}{\sqrt{-\z}}\llb \pa_a \(\sqrt{-\z} \z^{(ab)}\pa_b X^\m \) 
+ \ldots\rrb -\sum_i e_i \cN^{(i)}_a \sqrt{-\z} \z^{(ab)}\pa_b X^\m = 0  \; .
\lab{brane-X-eqs-a}
\ee
The dots in the l.h.s. of \rf{brane-X-eqs-a} indicate the same expression as
in the second line of Eq.\rf{brane-X-eqs} with $\P (\vp)$ substituted by 
$\sqrt{-\z}$.

Recalling again the definition \rf{def-N} of the normal $\cN^{(i)}_b$ w.r.t.
the charged $(p-1)$-sub-branes $\cB_i$ and the resulting property of the
dynamical tension $\cT\equiv \P (\vp)/\sqrt{-\z}$ being piece-wise constant
on the world-volume with jumps when crossing each sub-brane $\cB_i$, we find
that Eqs.\rf{brane-A-vec-eqs-a} imply:
\be
\pa_b {\wti H}^{ab} = 0 \quad ,\quad {\wti H}^{a\perp}\Bgv_{\cB_i} = 0 \; ,
\lab{brane-A-vec-eqs-b}
\ee
where the superscript $\perp$ indicates projection along the normal $N^{(i)}_a$
w.r.t. the world-hypersurface of sub-brane $\cB_i$. The first 
Eq.\rf{brane-A-vec-eqs-b} can be viewed as Bianchi identity for the field-strength
of a new $p-2$-form world-volume gauge field $B_{a_1\ldots a_{p-2}}$ (which
can be viewed as dual variable w.r.t. anti-symmetric part of $\z^{ab}$) :
\be
H_{a_1\ldots a_{p-1}} = (p-1)\pa_{[a_1} B_{a_2\ldots a_{p-1}]} \quad ,\quad
{\wti H}^{ab} \equiv \frac{1}{(p-1)!}
\vareps^{abc_1\ldots c_{p-1}} H_{c_1\ldots c_{p-1}}
\lab{H-eqs}
\ee
with ${\wti H}^{ab}$ as defined in \rf{brane-A-vec-eqs-a}. The second 
Eq.\rf{brane-A-vec-eqs-b} shows that the restrictions of the dual field strength
$H$ on each $(p-1)$-sub-brane $\cB_i$ vanish:
\be
H_{\a_1\ldots \a_{p-1}}\Bgv_{\cB_i} = 0
\lab{H-vanish-subbrane}
\ee
Here the $\a_1,\ldots ,\a_{p-1}$ indicate $p$-dimensional world-hypersurface 
tensorial indices on the sub-brane $\cB_i$. Let us note that the objects 
${\wti H}^{ab}$ and $B_{a_1\ldots a_{p-2}}$ already appeared in the
Polyakov-type formulation of the standard (non-modified) $Dp$-brane 
\ct{Hull-Zeid}. The new feature in the present modified-measure $Dp$-brane
model is the condition \rf{H-vanish-subbrane} on the dual field strength.

The same arguments from the previous paragraph applied to \rf{brane-X-eqs-a}
imply that $X^\m$ satisfy the standard $Dp$-brane equations of motion (in the
Polyakov-type formulation \ct{Hull-Zeid}) :
\br
\pa_a \(\sqrt{-\z} \z^{(ab)}\pa_b X^\m \) 
- \h \sqrt{-\z} G^{\m\n} \b \pa_\n U \,\z^{[ab]} F_{ab}(A) +
\nonu \\
\sqrt{-\z} \pa_a X^\l \pa_b X^\n \Bigl\lb \z^{(ab)} \G^\m_{\l\n}({\bar G}) + 
\h \z^{[ab]} G^{\m\k}\(\b \pa_\k U B_{\l\n} - \cH_{\k\l\n}(B)\) \Bigr\rb = 0  \; .
\lab{Dp-brane-X-eqs}
\er
together with Neumann boundary conditions on the world-hypersurfaces of each
charged $(p-1)$-sub-brane $\cB_i$ :
\be
\pa_{\perp} X^\m \Bgv_{\cB_i} = 0 \quad ,\quad
\pa_{\perp} \equiv N^{(i)}_a \z^{(ab)} \pa_b
\lab{brane-X-eqs-segments}
\ee
(recall that the symmetric part $\g^{ab} \equiv \z^{(ab)}$ plays the role of
(inverse) world-volume Riemannian metric).

Let us consider again Eq.\rf{brane-tension-eq} and integrate it along arbitrary
smooth closed curve $\G$ on the $Dp$-brane world-volume which is transversal to
(some or all of) the $(p-1)$-sub-brane $\cB_i$, we obtain the following 
constraints on the possible sub-brane configurations:
\be
\sum_i e_i\, n_i (\G) = 0 \; ,
\lab{charge-constr}
\ee
Here $n_i (\G)$ is the sign-weighted total number of $\G$ crossing $\cB_i$.
Eq.\rf{charge-constr} is the brane analog of the ``color'' charge confinement 
condition (first Eq.\rf{charge-constr-NA}) in the modified-measure string model.
In the present $p\geq 2$-brane case, however, due to the much more complicated
topologies of the pertinent world-volumes Eq.\rf{charge-constr} may yield various
different types of allowed sub-brane configurations. 

As a simple illustration, let us consider the simplest non-trivial case 
$p=2$ and take the static gauge for the $p=1$ sub-branes (strings), \textsl{i.e.},
the proper times of the charged strings coincides with the proper time of the
bulk membrane. The latter means that the fixed-time world-volume of the bulk
{\em closed} membrane is a Riemann surface with some number $g$ of handles
and no holes. Further, we will assume the following simple topology of the
attached $N$ charged strings $\cB_i$ : upon cutting the membrane surface
along these attached strings it splits into $N$ {\em open} membranes $\cM_i$
($i=1,\ldots ,N$) with Neumann boundary conditions (cf. \rf{brane-X-eqs-segments}), 
each of which being a Riemann surface with $g_i$ handles and
2 holes (boundaries) formed by the strings $\cB_{i-1}$ and $\cB_i$,
respectively\foot{The Euler characteristics of the bulk membrane Riemann
surface is $\chi = 2 -2g$, whereas for the open brane $\cM_i$ it is 
$\chi_i = 2 -2g_i -2$, so that $\chi = \sum_i \chi_i$ or, equivalently,
$g = 1 + \sum_i g_i$.}. The brane tension of $\cM_i$ is a dynamically
generated constant $T_i$ where $T_{i+1} = T_i + e_i$. In the present 
configuration Eq.\rf{charge-constr} evidently reduces to the
constraint $\sum_i e_i = 0$.

Thus, we conclude that similarly to the string case, modified-measure
$Dp$-brane models describe configurations of charged $(p-1)$-branes with
charge confinement. Apart from the latter, in general there exist more
complicated configurations allowed by the constraint \rf{charge-constr},
whose properties deserve further study.

\section{Conclusions}

Finally, let us sumarize the main features of the new class of modified-measure
string and brane models. The above discussion shows that:
\begin{itemize}
\item
There exist natural from physical point of view modifications of world-sheet and
world-volume integration measures which may significantly affect string and brane
dynamics. 
\item
Consistency of dynamics {\em naturally}~ requires the introduction of auxiliary
world-sheet vector gauge field (in the string case) and higher-rank world-volume
antisymmetric tensor gauge fields in the general brane case beyond the
standard $Dp$-brane $U(1)$ vector gauge field.
\item
The string/brane tension is {\em not}~ anymore a constant scale given {\em ad hoc},
but rather appears as an {\em additional dynamical degree of freedom}~ beyond the
ordinary string/brane degrees of freedom.
\item
The dynamical string/brane tension has physical meaning of an electric field
strength for the auxiliary world-sheet/world-volume gauge field.
\item
The dynamical string/brane tension obeys ``Gauss law'' constraint equation
and may be nontrivially variable in the presence of point-like charges (on
the string world-sheet) or charged lower-dimensional branes (on the
$Dp$-brane world-volume).
\item
Modified-measure string/brane models provide simple classical mechanisms for
confinement of point-like ``color'' charges or charged lower-dimensional branes
due to variable dynamical tension.
\end{itemize}

Possible applications of the new class of modified-measure $Dp$-brane models
in the context of modern brane-world theories are currently being studied.
As a step in this direction we refer to the recent paper \ct{conf-inv-bworld} 
where by employing the modified integration measure density \rf{D-measure-def}
a new conformally invariant brane-world model without (bulk) cosmological constant
fine tuning has been constructed. 

\textbf{Acknowledgements.} Two of us (E.N. and S.P.) are sincerely grateful
to Prof. Branko Dragovich for his cordial hospitality at the Second Summer
School on Modern Mathematical Physics, Kopaonik (Yugoslavia), 2002.
E.N. and S.P. are partially supported by Bulgarian NSF grant \textsl{F-904/99}.


\end{document}